\documentclass
[superscriptaddress,secnumarabic,nobibnotes,aps,prd,showkeys,noshowpacs,onecolumn,nofootinbib,10pt]{revtex4-2}%
\usepackage{graphics}
\usepackage{graphicx}
\usepackage{color}
\usepackage{epsf}
\usepackage{bm}
\usepackage{amsmath,amssymb,amsfonts,mathrsfs,amsthm}
\usepackage{latexsym}
\usepackage{enumerate}
\usepackage{comment}
\usepackage{hyperref}
\usepackage{epstopdf}
\usepackage{ulem}

\newcommand{\Lim}[1]{\raisebox{0.5ex}{\scalebox{0.8}{$\displaystyle \lim_{#1}\;$}}}

\begin{document}
\title{FLRW solutions in \texorpdfstring{$f(Q)$}{f(Q)} theory: the effect of using different connections}
\author{N. Dimakis}
\email{nsdimakis@scu.edu.cn}
\email{nsdimakis@gmail.com}
\affiliation{Center for Theoretical Physics, College of Physical Science and Technology, Sichuan University, Chengdu 610065, China}
\author{A. Paliathanasis}
\email{anpaliat@phys.uoa.gr}
\affiliation{Institute of Systems Science, Durban University of Technology,
PO Box 1334, Durban 4000, South Africa}
\affiliation{Instituto de Ciencias F\'isicas y Matem\'aticas,
Universidad Austral de Chile, Valdivia 5090000, Chile}
\author{M. Roumeliotis}
\email{microum@phys.uoa.gr}
\affiliation{Nuclear and Particle Physics section, Physics Department, National and Kapodistrian University of Athens, 15771 Athens, Greece}
\author{T. Christodoulakis}
\email{tchris@phys.uoa.gr}
\affiliation{Nuclear and Particle Physics section, Physics Department, National and Kapodistrian University of Athens, 15771 Athens, Greece}

\begin{abstract}
We study a Friedmann--Lema\^{\i}tre--Robertson--Walker (FLRW) space-time in the theory of $f(Q)$-gravity, where $Q$ denotes the non-metricity scalar. It has been previously shown in the literature, that there exist four distinct families of connections, which are compatible with the isometries of the FLRW metric; three for the spatially flat case and one when the spatial curvature is present. In the spatially flat case, one connection is dynamically irrelevant and yields the dynamics of the coincident gauge in the Cartesian coordinates. For this, we obtain the general solution of an arbitrary $f(Q)$ theory with a perfect fluid matter content, and present various examples for specific choices of the $f(Q)$ function. We proceed by studying the effect of the rest of the connections, which are dynamical and affect the equations of the motion. We concentrate in scenarios that depart from the $Q=$const. case, which just reproduces General Relativity with a cosmological constant, and derive novel vacuum solutions for a power-law $f(Q)$ function.
\end{abstract}

\maketitle

\section{Introduction}

Modified and extended theories of gravity \cite{clifton} play an important
role in the description of the cosmological observations \cite%
{Teg,Kowal,Komatsu,LHeis1}. In particular, the gravitational Action Integral is
modified by introducing geometric invariants. The result of this modification
is that new degrees of freedom are entering the gravitational field
equations so that to drive the dynamics away from that of General Relativity (GR) and thus provide a better theoretical
prediction for the observations \cite{cc1}. The Lagrangian density of Einstein's General Relativity is
based on the Ricci scalar, $R$, which is defined by the symmetric Levi-Civita connection.
The simplest modification of the Einstein-Hilbert Action is the introduction
of a function $f$, of this scalar. This leads to the
so-called $f( R) $-theory of gravity \cite{Buda}.

The Levi-Civita connection is not the unique choice which can be applied
in a gravitational theory. From a more general connection someone can define the
fundamental scalar invariants of the curvature, $R$, the torsion, $T$, and the
non-metricity, $Q$. The nature of the physical space depends on the invariant
which is used to define the gravitational Action Integral \cite{be1}.
Indeed, for the Levi-Civita connection, where only the curvature invariant
survives, the resulting theory is the General Relativity. On the other hand,
for the curvature-less Weitzenb{\"{o}}ck connection \cite{Weitzenb23}, we end up
with the Teleparallel Equivalent of General Relativity \cite%
{Hayashi79,Maluf:1994ji}. In addition, for a gravitational theory defined
by a torsion-free, flat connection we obtain the Symmetric Teleparallel
Equivalent of General Relativity \cite{Nester:1998mp}. While these three
theories admit the same field equations, this is not true for their
modifications. Indeed, the modified $f$-theories, for instance the $%
f\left( T\right) $-teleparallel theory \cite{f7} and the $f\left( Q\right) $%
-symmetric teleparallel theory \cite{f6} are quite distinct from the $f\left(
R\right) $-theory.

In this piece of study we are interested in the existence of cosmological
solutions for the $f\left( Q\right) $-symmetric teleparallel theory. There is a
plethora of studies in the literature on $f\left( Q\right) $-theory. Some
exact analytic solutions are presented in \cite{LHeis2,f15}. Power-law functions of $f(Q)$ and the values of the parameters that they admit have been studied in \cite{Capo}. The $f\left( Q\right) $-theory as a dark energy model is investigated in \cite{ff1,ff2,Chee,Lymperis}. A recent work on the properties of the effective fluid, owed to the non-metricity, can be found in \cite{Tsagas}. Some anisotropic spacetimes were studied in \cite{ww1,ww2,Esposito,Heis1,ww4,ww5}. The Hamiltonian analysis for the theory was performed in \cite{qq1}, while the quantization process in the case of cosmology was presented in \cite{qq2}. Wormhole solutions in the context of $f(Q)$ gravity are given in \cite{Banerjee}. Non-minimal couplings to matter have been considered in \cite{Harko2}, generalizations including the trace of the energy momentum tensor are found in \cite{Harko,Shiravand} and studies on observational constraints are  given in \cite{Ferreira,Ayuso}.

An important characteristic of $f\left( Q\right) $-theory is the use of a flat
connection pertaining the existence of affine coordinates in which all its
components vanish, turning covariant derivatives into partial (coincident
gauge). Thus, in $f(Q)$-theory it is possible to separate gravity from the inertial effects.
Hence, the construction of the $f\left( Q\right) $-theory forms a novel
starting point for various modified gravity theories. It also presents a
simple formulation in which self-accelerating solutions arise naturally in
both the early and late Universe.

Although the coincident gauge is always achievable through an appropriate coordinate transformation, extra care is needed when we a priori adopt a specific coordinate system through an ansatz for the space-time metric \cite{Zhao}. This is because, when starting from a particular line element, e.g. a Friedmann--Lema\^{\i}tre--Robertson--Walker (FLRW) metric expressed in spherical coordinates, we already have partially fixed the gauge. Thus, the connection may not become zero at the coordinate system which we have already assumed for the metric. In particular for the FLRW case, it has been shown \cite{Heis1,Hohmann}, that there exist four distinct possible connections that are compatible with its isometries; three for the spatially flat case and one when the spatial curvature is not zero. One of the three connections of the spatially flat case, becomes zero, when we transform the metric to Cartesian coordinates; the rest assume their coincident gauge form in totally different coordinate systems, which lead to metrics having non-diagonal terms. In this work we derive the field equations in FLRW geometry, with or without spatial curvature, for all four existing forms of the symmetric, flat
connection. We prove the existence of interesting analytic solutions in all of the cases involving the different connections and study specific examples.

The plan of the paper is as follows: In Section \ref{sec2} we present the basic properties and definitions of $f\left(
Q\right) $-theory. In Section \ref{sec3}, we present the symmetric connection which
are compatible with the FLRW geometry in spherical coordinates. For the spatially flat FLRW geometry, the field
equations are derived in Section \ref{sec4}. Exact solutions are determined for the
latter and we show that inflationary solutions exist. The case of a
non-zero spatial curvature is assumed in Section \ref{sec5}. Finally, in Section \ref{sec6} we
draw our conclusions.

\section{Preliminaries} \label{sec2}

In metric-affine gravitational theories, the basic dynamical objects are the metric $g_{\mu\nu}$ and the connection $\Gamma^\kappa_{\; \mu\nu}$. The fundamental tensors that can be constructed with the help of these objects are the  curvature $R^\kappa_{\;\lambda\mu\nu}$, the torsion $\mathrm{T}^\lambda_{\mu\nu}$ and the non-metricity $Q_{\lambda\mu\nu}$, whose components are given respectively by
\begin{align}
   R^\kappa_{\;\lambda\mu\nu} & = \frac{\partial \Gamma^\kappa_{\;\lambda \nu}}{\partial x^\mu} - \frac{\partial \Gamma^\kappa_{\;\lambda \mu}}{\partial x^\nu} + \Gamma^\sigma_{\; \lambda \nu} \Gamma^\kappa_{\; \mu\sigma} - \Gamma^\sigma_{\; \lambda \mu} \Gamma^\kappa_{\; \mu \sigma} \\
   \mathrm{T}^\lambda_{\mu\nu} & = \Gamma^{\lambda}_{\; \mu\nu} - \Gamma^{\lambda}_{\; \nu\mu} \\
  Q_{\lambda\mu\nu} &  = \nabla_{\lambda} g_{\mu\nu} = \frac{\partial g_{\mu\nu}}{\partial x^\lambda} - \Gamma^\sigma_{\; \lambda\mu} g_{\sigma\nu} - \Gamma^\sigma_{\; \lambda\nu} g_{\mu\sigma}.
\end{align}
In the above relations $\nabla_\mu$ is used to denote the covariant derivative with respect to the affine connection $\Gamma^\kappa_{\; \mu\nu}$ and $x^\mu$ are the coordinate components upon the manifold. In the case of a symmetric connection the torsion is zero, $\mathrm{T}^\lambda_{\mu\nu}=0$. This, together with the condition $Q_{\lambda\mu\nu}=0$, results in the well-known metric theories of gravity.

In an attempt to consider theories outside the scope of (pseudo-)Riemannian geometry, more general connections are taken into account, which lead to the torsion and/or the non-metricity being non-zero. In the theory of symmetric teleparallelism and its modifications, in which we are interested in this work, the flatness, $R^\kappa_{\;\lambda\mu\nu}=0$ and the torsionless, $\mathrm{T}^\lambda_{\mu\nu}=0$, conditions are imposed, leaving only $Q_{\lambda\mu\nu}\neq 0$. The basic geometric scalar of the theory is defined as
\begin{equation} \label{defQ}
  Q= Q_{\lambda\mu\nu} P^{\lambda\mu\nu},
\end{equation}
where $P^\lambda_{\;\mu\nu}$ is used for the components of the non-metricity conjugate tensor
\begin{equation} \label{defP}
  P^\lambda_{\;\mu\nu}= - \frac{1}{4} Q^{\lambda}_{\; \mu\nu} + \frac{1}{2} Q^{\phantom{(\mu} \lambda \phantom{\nu)}}_{(\mu \phantom{\lambda} \nu)} + \frac{1}{4} \left(Q^\lambda- \bar{Q}^\lambda\right) g_{\mu\nu} -\frac{1}{4} \delta^{\lambda}_{\; (\mu}Q_{\nu)},
\end{equation}
which is written with the help of the traces $Q_\mu= Q_{\mu \nu}^{\phantom{\mu\nu} \nu}$ and $\bar{Q}_\mu= Q_{\phantom{\nu} \mu \nu}^{\nu \phantom{\mu} \phantom{\mu}}$. The $\delta^{\mu}_{\;\nu}$ in \eqref{defP} is the Kroncker delta and the parentheses in the indices denote the usual symmetrization, i.e. $A_{(\mu\nu)}=\frac{1}{2} \left(A_{\mu\nu}+A_{\nu\mu}\right)$.

The non-metricity scalar $Q$ of \eqref{defQ} is defined in such a way so that, when taken as a Lagrangian density, the theory which it produces, is dynamically equivalent to Einstein's general relativity. Non-linear generalizations of symmetric teleparallelism, involve a gravitational Lagrangian density which is characterized by a generally non-linear function $f(Q)$
\begin{equation}\label{action}
  S = \frac{1}{2} \int d^4x  \sqrt{-g} f(Q) + \int d^4x \sqrt{-g} \mathcal{L}_M + \lambda_{\kappa}^{\; \lambda\mu\nu} R^{\kappa}_{\; \lambda\mu\nu} + \tau_{\lambda}^{\; \mu\nu} \mathrm{T}^\lambda_{\;\mu\nu}.
\end{equation}
In the above action, $g=\mathrm{det}(g_{\mu\nu})$, is the determinant of the space-time metric, the $\mathcal{L}_M$ is the matter fields' Lagrangian density, while $\lambda_{\kappa}^{\; \lambda\mu\nu}$ and $\tau_{\lambda}^{\; \mu\nu}$ are Lagrange multipliers, whose variation enforces the flatness and torsionless conditions $R^{\kappa}_{\; \lambda\mu\nu}=0=\mathrm{T}^\lambda_{\;\mu\nu}$. The above action, and the subsequent results in this work, are expressed in units $8\pi G=c=1$.

Variation of \eqref{action} with respect to the metric results in \cite{Harko}
\begin{equation}\label{feq1a}
  \frac{2}{\sqrt{-g}} \nabla_{\lambda}\left(\sqrt{-g} f'(Q) P^\lambda_{\; \mu\nu} \right) - \frac{1}{2}f(Q) g_{\mu\nu} + f'(Q) \left(P_{\mu\rho\sigma}Q_{\nu}^{\;\rho\sigma}- 2 Q_{\rho\sigma\mu}P^{\rho\sigma}_{\phantom{\rho\sigma}\nu}\right) = T_{\mu\nu},
\end{equation}
where the primes are used to express derivation with respect to the argument, e.g. $f'(Q)=\frac{df}{dQ}$. The $T_{\mu\nu}=-\frac{2}{\sqrt{-g}} \frac{\partial \left(\sqrt{-g}\mathcal{L}_M\right)}{\partial g^{\mu\nu}}$ is the energy-momentum tensor emerging from the matter contribution in the action \eqref{action}.  Variation with respect to the connection leads to the additional equations of motion
\begin{equation}\label{feq2}
  \nabla_{\mu}\nabla_{\nu} \left( \sqrt{-g} f'(Q) P^{\mu\nu}_{\phantom{\mu\nu}\sigma} \right) =0 .
\end{equation}
The set of equations \eqref{feq1a} can assume the more convenient expression \cite{Zhao}
\begin{equation} \label{feq1}
   f'(Q) G_{\mu\nu} + \frac{1}{2} g_{\mu\nu} \left( f'(Q) Q- f(Q) \right) + 2 f''(Q) \left(\nabla_{\lambda}Q\right) P^\lambda_{\; \mu\nu} = T_{\mu\nu},
\end{equation}
where $G_{\mu\nu}= \tilde{R}_{\mu\nu}-\frac{1}{2}g_{\mu\nu} \tilde{R}$ is the usual Einstein tensor, with $\tilde{R}_{\mu\nu}$ and $\tilde{R}$ being the Riemannian Ricci tensor and scalar respectively (constructed with the Levi-Civita connection). In this form the equations for the metric allow a direct comparison with General Relativity (GR) since the (dynamical) deviation from the latter can be perceived as the effect of an effective energy momentum tensor
\begin{equation}\label{Teff}
  \mathcal{T}_{\mu\nu} = -\frac{1}{f'(Q)} \left[ \frac{1}{2} g_{\mu\nu} \left( f'(Q) Q- f(Q) \right) + 2 f''(Q) \left(\nabla_{\lambda}Q \right) P^\lambda_{\; \mu\nu} \right] .
\end{equation}
With the help of \eqref{Teff}, equation \eqref{feq1} becomes $G_{\mu\nu} = \mathcal{T}_{\mu\nu} + \frac{1}{f'(Q)}T_{\mu\nu}$, which reveals the role of $\mathcal{T}_{\mu\nu}$ as that of an energy-momentum tensor of geometric origin. From \eqref{Teff} it can be directly seen that $f(Q)\propto Q$ results in the same equations as GR, since the assumption leads to $\mathcal{T}_{\mu\nu}=0$. It is also obvious, that the case $Q=$const. leads to solutions of GR plus a cosmological constant $\Lambda$ whose value is $\Lambda = \frac{1}{2}\left(Q- \frac{f(Q)}{f'(Q)}\right)$.

As is well-known \cite{Eisenhart}, the flatness condition, $R^\kappa_{\;\lambda\mu\nu}=0$, implies that there exists a coordinate system in which the connection becomes zero, $\Gamma^\lambda_{\;\mu\nu}=0$. This is usually referred to as the coincident gauge. However, special care is needed when the equations of motion are  considered after a partial gauge fixing at the level of the metric. For example, when we take a FLRW space-time or a static and spherically symmetric manifold, there is the possibility that the gauge in which $\Gamma^\lambda_{\;\mu\nu}=0$ is realised is incompatible with the coordinate system in which the metric is expressed and this may lead to unnecessary restrictions in the equations of motion \cite{Zhao,Heis1}.

Another interesting point, mentioned in \cite{Eisenhart}, is that all flat spaces are necessarily Riemannian; this however, in the sense that, if a connection $\Gamma^\lambda_{\;\mu\nu}$ is flat, leading to $R^\kappa_{\;\lambda\mu\nu}=0$, then there must exist some metric $\bar{g}_{\mu\nu}$ for which the $\Gamma^\lambda_{\;\mu\nu}$ are its Christoffel symbols. Of course, in our case, the $g_{\mu\nu}$ we consider is a completely disassociated object from this $\bar{g}_{\mu\nu}$ and independent from the connection, thus allowing us to have $Q_{\lambda\mu\nu}\neq 0$.

In symmetric teleparallel gravity and its modifications, for a matter content which is minimally coupled to the metric, the conservation law $T^{\mu}_{\phantom{\mu}\nu;\mu}=0$ holds for the matter energy-momentum tensor. The semicolon ``$;$'' here is used to denote the covariant derivative with respect to the Christoffel symbols. The $T^{\mu}_{\phantom{\mu}\nu;\mu}=0$ relation holds by virtue of equation \eqref{feq2} for the connection, which in itself it can also be perceived as a conservation law for the theory \cite{Koiv}.

\section{FLRW space-time} \label{sec3}

We start by writing the FLRW line element, which in spherical coordinates reads
\begin{equation} \label{genlineel}
  ds^2 = - N(t)^2 dt^2 + a(t)^2 \left[ \frac{dr^2}{1-k r^2} + r^2 \left( d\theta^2 + \sin^2\theta d\phi^2 \right) \right] .
\end{equation}
In \cite{Hohmann,Heis2} the general form of all compatible connections with \eqref{genlineel}, that lead to a zero curvature tensor, has been derived. This has been done by enforcing on a generic connection the six Killing symmetries of \eqref{genlineel}, plus the demand to satisfy $R^\kappa_{\;\lambda\mu\nu}=0$.

The six isometries of \eqref{genlineel}, associated with the isotropy and the homogeneity of space, are given in the above coordinates by
\begin{equation} \label{Kil1}
  \zeta_1 = \sin\phi \partial_\theta + \frac{\cos\phi}{\tan\theta} \partial_\phi, \quad \zeta_2 = -\cos\phi \partial_\theta + \frac{\sin\phi}{\tan\theta} \partial_\phi, \quad  \zeta_3 = - \partial_\phi
\end{equation}
and
\begin{equation} \label{Kil2}
  \begin{split}
    \xi_1 & = \sqrt{1-k r^2}\sin\theta \cos\phi \partial_r + \frac{\sqrt{1-k r^2}}{r} \cos\theta \cos\phi \partial_\theta - \frac{\sqrt{1-k r^2}}{r} \frac{\sin\phi}{\sin\theta} \partial_\phi \\
    \xi_2 & = \sqrt{1-k r^2}\sin\theta \sin\phi \partial_r + \frac{\sqrt{1-k r^2}}{r} \cos\theta \sin\phi \partial_\theta + \frac{\sqrt{1-k r^2}}{r} \frac{\cos\phi}{\sin\theta} \partial_\phi \\
    \xi_3 & = \sqrt{1-k r^2} \cos\theta \partial_r - \frac{\sqrt{1-k r^2}}{r} \sin\theta \partial_\phi .
  \end{split}
\end{equation}
The Lie derivative of an affine connection with respect to a vector $X$ is calculated to be \cite{Bardeen}
\begin{equation}
  \mathcal{L}_X \Gamma^\mu_{\;\kappa\lambda} = X^\sigma \frac{\partial \Gamma^\mu_{\;\kappa\lambda}}{\partial x^\sigma}  + \Gamma^\mu_{\;\sigma\lambda} \frac{\partial X^\sigma}{\partial x^\kappa} + \Gamma^\mu_{\;\kappa\sigma} \frac{\partial X^\sigma}{\partial x^\lambda} - \Gamma^\sigma_{\;\kappa\lambda} \frac{\partial X^\mu}{\partial x^\sigma} + \frac{\partial^2 X^\mu}{\partial x^\kappa \partial x^\lambda} .
\end{equation}
The requirement $R^\kappa_{\;\lambda\mu\nu}=0$, in conjunction with $\mathcal{L}_X \Gamma^\mu_{\;\kappa\lambda}=0$, where $X$ is any of the $\zeta_i$ or $\xi_i$ ($i=1,2,3$), lead to the following possibilities:

\begin{itemize}
  \item \textbf{Spatially flat case $k=0$}. There are three admissible connections. The common non-zero components that all three have are the following:
  \begin{equation} \label{common}
    \begin{split}
     & \Gamma^r_{\;\theta\theta}=-r, \quad \Gamma^r_{\;\phi\phi}=-r \sin^2\theta, \\
     & \Gamma^\theta_{\; r\theta}= \Gamma^\theta_{\;\theta r}=\Gamma^\phi_{\; r\phi} =\Gamma^\phi_{\;\phi r}= \frac{1}{r}, \quad \Gamma^\theta_{\;\phi\phi}=-\sin\theta \cos\theta, \Gamma^\phi_{\; \theta\phi}=\Gamma^\phi_{\; \phi\theta} = \cot\theta.
    \end{split}
  \end{equation}
  However, they do differ in the way a free function of time enters in some of their other components. The first connection has only one additional non-zero component
  \begin{equation} \label{con1}
    \Gamma^t_{\;tt} = \gamma(t),
  \end{equation}
  where $\gamma(t)$ is a function of the time variable $t$. The second connection has the following extra non-zero components
  \begin{equation} \label{con2}
      \Gamma^t_{\;tt} = \frac{\dot{\gamma}(t)}{\gamma(t)} + \gamma(t), \quad \Gamma^r_{\;tr}= \Gamma^r_{\;rt} =\Gamma^\theta_{\;t\theta}= \Gamma^\theta_{\;\theta t}= \Gamma^\phi_{\;t\phi}= \Gamma^\phi_{\;\phi t} =\gamma(t),
  \end{equation}
  where the dot denotes differentiation with respect to $t$. Finally, the third connection has the additional components
  \begin{equation} \label{con3}
      \Gamma^t_{\;tt} = -\frac{\dot{\gamma}(t)}{\gamma(t)}, \quad \Gamma^t_{\;rr} = \gamma(t), \quad \Gamma^t_{\;\theta\theta} = \gamma(t) r^2, \quad \Gamma^t_{\;\phi\phi} = \gamma(t) r^2 \sin^2\theta .
  \end{equation}
  The first connection, consisting of \eqref{common} and \eqref{con1}, when $\gamma(t)=0$, becomes itself zero when transforming it from spherical to Cartesian coordinates. In other words, it corresponds to the coincident gauge in the latter coordinate system \cite{Zhao}.

  \item \textbf{Case of non-zero spatial curvature $k\neq0$}. Here, the following connection is obtained (listing again only the non-zero components):
  \begin{equation}\label{conk1}
    \begin{split}
      & \Gamma^t_{\;tt} = - \frac{k+\dot{\gamma}(t)}{\gamma (t)}, \quad \Gamma^t_{\;rr} = \frac{\gamma (t)}{1-k r^2} \quad \Gamma^t_{\;\theta\theta} = \gamma(t) r^2 , \quad \Gamma^t_{\;\phi\phi} = \gamma (t) r^2 \sin ^2(\theta ) \\
      & \Gamma^r_{\;tr}= \Gamma^r_{\;rt}= \Gamma^\theta_{\;t\theta}=\Gamma^\theta_{\;\theta t} =\Gamma^\phi_{\;t\phi}= \Gamma^\phi_{\;\phi t} = -\frac{k}{\gamma(t)}, \quad \Gamma^r_{\;rr} = \frac{k r}{1-k r^2}, \quad \Gamma^r_{\;\theta\theta} = -r \left(1- k r^2\right),  \\
      & \Gamma^r_{\;\phi\phi} = - r \sin ^2(\theta ) \left(1-k r^2\right) \quad \Gamma^\theta_{\;r\theta}=\Gamma^\theta_{\;\theta r} = \Gamma^\phi_{\;r\phi} = \Gamma^\phi_{\;\phi r} = \frac{1}{r}, \quad \Gamma^\theta_{\;\phi\phi} = -\sin\theta\cos\theta, \quad \Gamma^\phi_{\;\theta\phi} =\Gamma^\phi_{\;\phi\theta} = \cot\theta .
    \end{split}
  \end{equation}
  As it is obvious, when $k=0$, the above connection yields the third one from the previous set consisting of \eqref{common} and \eqref{con3}. This connection has also been presented previously in various works \cite{Zhao,Hohmann,Heis2}.\footnote{For the convenience of the reader, and to avoid possible confusion, we just mention that there is a minor typo in the expression of the connection \eqref{conk1} as is given in \cite{Heis2}. The authors there use $\chi=\sqrt{1-k r^2}$ inside the connection, when actually, as we see from \eqref{conk1}, it should be $\chi^2=1-k r^2$.}
\end{itemize}

\section{Spatially flat case} \label{sec4}

It is usual in the literature of $f(Q)$ theory to study the cosmological aspects of a spatially flat FLRW line element in Cartesian coordinates $ds^2= -N^2 dt^2 +a(t)^2 (dx^2+dy^2+dz^2)$ and in the coincident gauge $\Gamma^{\mu}_{\;\kappa\lambda}=0$. In the spherical coordinates, where the line element is given by \eqref{genlineel}, this corresponds to taking the first connection with $\gamma=0$. Here, we are interested to see how all possible connections may affect the dynamics and compare the obtained solutions to what happens in General Relativity.

\subsection{First connection}

We start by considering the connection $\Gamma^{\mu}_{\;\kappa\lambda}$ whose non-zero components are given by \eqref{common} and \eqref{con1}. The emerging dynamics is equivalent to that of the coincident gauge, since the function $\gamma$ does not appear in the resulting expression for $Q$. The latter is obtained, from the definition \eqref{defQ}, to be
\begin{equation}\label{Qcon1}
  Q = - \frac{6\dot{a}^2}{N^2 a^2} = - 6 H^2.
\end{equation}
In the above relation we have used $H = \frac{\dot{a}}{N a}$ for the Hubble function as expressed in the time gauge where the lapse is $N(t)$. When we are at the cosmic time gauge, $N=1$, of course we obtain the well-known $H=\frac{\dot{a}}{a}$. At this point however, we shall avoid fixing the gauge, since we are going to utilize this freedom later on, in order to simplify the process of obtaining solutions from the field equations. Another point that needs to be made is, that in our conventions, the non-metricity scalar $Q$, as can be seen by \eqref{Qcon1}, is negative (or possibly zero). This is because we defined it as $Q=Q_{\lambda\mu\nu} P^{\lambda\mu\nu}$. There are works in the literature where $Q$ is taken as $Q=-Q_{\lambda\mu\nu} P^{\lambda\mu\nu}$, which yields a positive $Q$. This is similar to what happens in General Relativity where there exist two equivalent definitions for the Riemann curvature differing only in an overall sign. We mention this point so that it can be taken into account when comparing with other works in the literature using a different convention and thus avoid any confusion.

The equations of motion for the connection, \eqref{feq2}, are identically satisfied, while those for the metric, Eqs. \eqref{feq1}, are independent of $\gamma(t)$ and they are equivalent to
\begin{subequations} \label{feq11}
  \begin{align} \label{feq11a}
    & \frac{3 \dot{a}^2}{N^2 a^2} f'(Q) + \frac{1}{2} \left( f(Q) - Q f'(Q) \right) = \rho \\ \label{feq11b}
    & -\frac{2}{N} \frac{d}{dt} \left( \frac{f'(Q) \dot{a}}{N a} \right) - \frac{3 \dot{a}^2}{N^2 a^2} f'(Q) - \frac{1}{2} \left(f(Q)- Q f'(Q)\right) = p,
  \end{align}
\end{subequations}
where we have used $T^{\mu}_{\phantom{\mu}\nu}= \mathrm{diag}(-\rho,p,p,p)$ for the energy momentum tensor. Note that in the above equation, we have not substituted $Q$ from its expression given in \eqref{Qcon1}. For consistency, it is easy to see that upon setting $f(Q)=Q-2\Lambda$, and considering the cosmic time gauge $N=1$, the above equations reduce to the well known Friedmann equations (remember we work in units $8\pi G=c=1$)
\begin{align}
  & \frac{\dot{a}^2}{a^2} = \frac{1}{3} \left( \rho + \Lambda \right) \\
  & \frac{\ddot{a}}{a} = \frac{\Lambda}{3} - \frac{1}{2} \left(p+\frac{\rho}{3} \right) .
\end{align}
Thus, General Relativity is recovered for a linear $f(Q)$ function, as it is expected. But, let us return to the generic problem described by \eqref{feq11}. As a first observation it is easy to see that in the case of vacuum, $p=\rho=0$, the two equations combined result in the constraint
\begin{equation}
   \left(f(Q)-2 Q f'(Q)\right) \dot{Q} f''(Q) =0 .
\end{equation}
This is obtained if you solve \eqref{feq11a} algebraically with respect to the lapse and substitute the latter in \eqref{feq11b}. From the above relation we distinguish three possibilities: i) First, we get a theory with $f(Q)\propto \sqrt{-Q}$, for which all equations are identically satisfied. If we would write the minisuperspace Lagrangian which produces $\eqref{feq11}$ as Euler-Lagrange equations, we would see that $f(Q)\propto \sqrt{-Q}$ turns the Lagrangian into a total derivative, i.e. the action is a pure surface term and as a result the equations are trivially satisfied. ii) The non-metricity scalar is constant, $Q=$const., which, as previously mentioned, is equivalent to having General Relativity with a cosmological constant. This leads to the known de-Sitter solution, $N=1$, $a=e^{\pm \sqrt{\frac{\Lambda}{3}}t}$, with $\Lambda$ acquiring the value $\Lambda = \frac{1}{2}\left(Q- \frac{f(Q)}{f'(Q)}\right)$. iii) Finally, there is the possibility that $f(Q)$ is a linear in $Q$ function which again yields the relativistic solutions (either de-Sitter or the flat space). So, we see that in the context of the first connection there are no vacuum solutions outside General Relativity, not unless we consider $f(Q)\propto \sqrt{-Q}$, which however gives rises to infinitely many solutions thus stripping the theory of any predictability.

The situation changes with the consideration of matter. Let us consider a perfect fluid of the typical linear barotropic equation $p=w \rho$. It has been shown that the equations give rise to the same continuity equation as the one emerging in metric theories of gravity \cite{Harko}
\begin{equation}
  \dot{\rho} +\frac{3\dot{a}}{a} \left( \rho+p \right) =0.
\end{equation}
The above leads the well-known solution $\rho=\rho_0 a^{-3(1+w)}$, with $\rho_0$ being the integration constant. By solving equation \eqref{feq11a} with respect to the lapse, we obtain
\begin{equation} \label{sol1N}
  N(t) = \pm \frac{\dot{a}}{a} \left( \frac{6 f'(Q)}{2\rho+ Q f'(Q)-f(Q)} \right)^{\frac{1}{2}}.
\end{equation}
With its substitution in the second equation, \eqref{feq11b}, the latter becomes
\begin{equation}
  \left(2 \rho_0 - a^{3(1+ w)} \left(f(Q)-2 Q f'(Q)\right)\right) \dot{Q} f''(Q)=0 .
\end{equation}
Assuming that we want to encounter solutions that are distinguishable from General Relativity, we need to consider $\dot{Q}\neq0 $ and $f''(Q)\neq 0$ (later we are going to see what happens if a posteriori we set $f(Q)=Q$ in our result). The above relation can be simply solved algebraically with respect to the scale factor (as long as $w\neq-1$) leading to
\begin{equation} \label{sol1a}
  a(t) = \left( \frac{2 \rho_0}{f(Q)-2 Q f'(Q)}\right)^{\frac{1}{3(1+ w)}}.
\end{equation}
Of course we consider that $f(Q)$ is not proportional to $\sqrt{-Q}$, so the denominator cannot be zero. Up to now we have not made use of the freedom of fixing the time gauge. We may choose to set $Q=-t$ which will give us straightforwardly $N(t)$ and $a(t)$ from \eqref{sol1N} and \eqref{sol1a} for any given $f(Q)$ theory. We insert the minus sign for simplicity, because, as we already mentioned, in our definition $Q$ is negative, thus, by setting $Q=-t$ the solution will be valid in the positive half line $t\in\mathbb{R}_+$ (if we had set $Q=t$ we would need to consider $t\in\mathbb{R}_{-}$). We thus see that, in the chosen time gauge, we are able to express the solution in terms of elementary functions, assuming of course that $f(Q)$ is also such a function.

In order to check the validity of the result, by testing if it can be connected to a relativistic solution, let us consider setting $f(Q)=Q$ into \eqref{sol1N} and \eqref{sol1a}. At the same time let us make the gauge fixing choice $Q=-t$. With these substitutions, equations \eqref{sol1N} and \eqref{sol1a} result in
\begin{align} \label{ex1N}
   N(t) = \pm \sqrt{\frac{2}{3}} \frac{1}{(1+w)t^{\frac{3}{2}}} \\
   a(t) = \left(\frac{2 \rho_0}{t}\right)^{\frac{1}{3(1+w)}}.
\end{align}
We may recognize this solution if we transform it into the cosmic time gauge, where $N(\tau)=1$. From now on we shall use $\tau$ to denote the time in that gauge. We thus want to make a mapping $t\rightarrow \tau$ that yields $N(\tau)=1$. From the transformation law of the lapse function, we have
\begin{equation} \label{gentimetr}
   N(t)dt = N(\tau) d\tau \Rightarrow \int\!\! N(t)dt = \tau + C .
\end{equation}
For the $N(t)$ given by \eqref{ex1N}, we get
\begin{equation}
  \sqrt{\frac{2}{3}} \frac{1}{(1+w)t^{\frac{3}{2}}} dt = d\tau \Rightarrow \sqrt{\frac{2}{3}} \frac{1}{(1+w)}\int\!\! \frac{1}{t^{\frac{3}{2}}} dt = \tau ,
\end{equation}
where, in order to simplify our considerations, we just chose to use the positive branch of \eqref{ex1N} and set the integration constant, $C$, on the right hand side of the expression \eqref{gentimetr} equal to zero. If we solve to find $t$ as a function of $\tau$ we obtain
\begin{equation}
  t= \frac{8}{3\rho_0 (1+w)^2 \tau^2}.
\end{equation}
With the use of this mapping from $t$ to $\tau$ we are led to $N(\tau)=1$, while the scale factor becomes
\begin{equation}
  a(\tau) = \left(\frac{3 \rho_0(1+w)^2}{4}\right)^{\frac{1}{3(1+w)}} \tau^{\frac{2}{3(1+w)}}.
\end{equation}
This is none other than the well-known perfect fluid solution of Einstein's equations $G_{\mu\nu}=T_{\mu\nu}$. Hence, we see that \eqref{sol1N} and \eqref{sol1a} truly retrieve the General Relativistic solution when $f(Q)=Q$.

We now gather the pair that forms the solution
\begin{subequations} \label{fsol1}
  \begin{align}
    N(t) & = \pm \left(-\frac{2}{3 Q}\right)^{\frac{1}{2}} \frac{\left(2 Q f''(Q)+f'(Q)\right) \dot{Q}}{(w+1)\left(f(Q)-2 Q f'(Q)\right)} \\
    a(t) & = \left( \frac{2 \rho_0}{f(Q)-2 Q f'(Q)}\right)^{\frac{1}{3(1+ w)}} ,
  \end{align}
\end{subequations}
which we can use to write the general line element that solves the equations, with a perfect fluid satisfying a linear barotropic equation, for any $f(Q)$ theory with a non-constant $Q$,
\begin{equation} \label{lineel1}
  ds^2 = \frac{2 \left(2 Q f''(Q)+f'(Q)\right)^2}{3 (w+1)^2 Q \left(f(Q)-2 Q f'(Q)\right)^2} dQ^2 + \left( \frac{2 \rho_0}{f(Q)-2 Q f'(Q)}\right)^{\frac{2}{3(1+ w)}} \left(dr^2 + r^2 \left(d\theta^2 + \sin^2\theta d\phi^2\right)\right) .
\end{equation}
The $-Q$ assumes effectively the role of the time variable. Note that the solution is of Lorentzian signature for $Q<0$, which is consistent with the convention we use.

Of course we need to note that although we arrive at a specific metric that forms a solution, there exists a degeneracy due to the fact that the function entering the connection remains arbitrary. Thus, we are dealing in reality with infinitely many solutions, one for each distinct $\gamma(t)$ function.

We also need to mention that the above solution can also be obtained through studying the inverse problem, i.e. the derivation of the matter content with respect to the gravitational functions. Equations \eqref{feq11} can obviously be perceived as definitions for $\rho$ and $p$. Then, we need only observe that we can integrate \eqref{Qcon1} with respect to $a(t)$. This supplements as with $a$ as an integral $N$ and $Q$:
\begin{equation} \label{tchrisgen}
  a(t) = a_0 \exp\left(\int\!\! N\sqrt{-Q} dt \right).
\end{equation}
At this point, equations \eqref{feq11} can be considered solved (having given the necessary $\rho$ and $p$) and the corresponding scale factor is given just by \eqref{tchrisgen}. The $-Q$ plays again the role of the time variable, while $N$ is arbitrary. The latter will obtain a particular dependence if we decide to adopt some specific equation of state. For example, if we set $p=w\rho$ with $\rho$ and $p$ given by \eqref{feq11} with the substitution of \eqref{tchrisgen}, then the $p=w\rho$ is algebraically solvable with respect to $N$, yielding again solution \eqref{fsol1}. One needs not be restricted to setting $p=w\rho$, different equations of state can also be assumed, leading to distinct results. However, in this work, we just restrict our attention to the linear equation of state.

At this point it is useful to consider some examples and see how \eqref{lineel1} can be used to derive conclusions about the evolution implied by some specific $f(Q)$ functions.

\subsubsection{The \texorpdfstring{$f(Q)=Q + \alpha Q^\mu$}{f(Q)=Q+a Q\^{}m} example}

From relations \eqref{fsol1}, or equivalently from \eqref{lineel1}, we may write the lapse and the scale factor, for a function $f(Q)=Q + \alpha Q^\mu$, in the time gauge $Q=-t$, as
\begin{subequations} \label{solex1}
  \begin{align} \label{solex1N}
    N = & \pm \sqrt{\frac{2}{3}} \frac{t-\alpha  \mu  (2 \mu -1) (-t)^{\mu }}{t^{3/2} (w+1) \left(t-\alpha  (2 \mu -1) (-t)^{\mu }\right)} \\
    a = & \left[\frac{2 \rho_0}{t-\alpha (2 \mu -1) (-t)^{\mu }}\right]^{\frac{1}{3( w+1)}}.
  \end{align}
\end{subequations}
The corresponding energy density for the perfect fluid is given by
\begin{equation} \label{solex1rho}
  \rho=\rho_0 a^{-3(1+w)}= \frac{1}{2} \left(t-\alpha   (2 \mu -1) (-t)^{\mu }\right)
\end{equation}
For simplicity, we shall consider integer values for $\mu$, so that $\alpha$ is restricted to be real in order to have a real valued $f(Q)=Q + \alpha Q^\mu$ function when $Q<0$. With this consideration, we may observe from \eqref{solex1}, that, if we want a solution of Lorentzian signature, we need to set the restriction $t>0$. The expression $t-\alpha (2 \mu -1) (-t)^{\mu } $ that we see in the scale factor can be either positive or negative with the additional condition that the constant $\rho_0$ must also be of the same sign. By \eqref{solex1rho} however, we see that considering $t-\alpha (2 \mu -1) (-t)^{\mu } <0$  leads to a negative energy density $\rho$. Although there exist matter contents that give rise to such a negative energy density  \cite{Nemiroff}, for our example, let us consider the more usual case where $\rho>0$. Thus, in the end,  we require
\begin{equation} \label{exLor}
  t>0 \quad \text{and} \quad t-\alpha (2 \mu -1) (-t)^{\mu } >0.
\end{equation}
For some particular values of $\mu$ the behaviour of the Hubble function, $H$ with respect to the scale factor can be extracted directly by solving algebraically the temporal field equation with respect to $H$. For example, for $\mu=2$ the equation \eqref{feq11a} leads to
\begin{equation}
   54 \alpha  H^4 - 3 H^2 + \rho_0 a^{-3 (w+1)} =0 .
\end{equation}
The latter is of course algebraically solvable with respect to $H$.

\begin{figure}[ptb]
\includegraphics[width=1\textwidth]{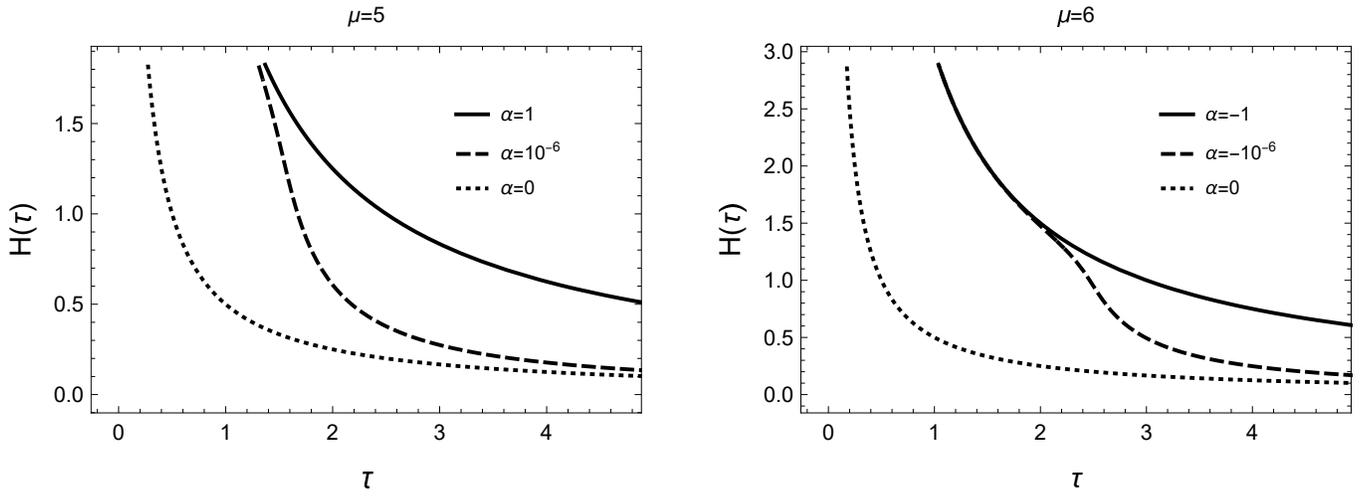}\caption{Plots of the Hubble function in $f(Q)=Q + \alpha Q^\mu$ theory for $\mu=5$ and $\mu=6$ and for different orders of magnitude of the coupling constant $\alpha$ in the cosmic time, $\tau$, gauge. In both graphs the corresponding GR solution is displayed with the dotted $\alpha=0$ line. For the equation of state parameter we have considered $w=\frac{1}{3}$.}%
\label{fig1}%
\end{figure}

In order to make use of \eqref{solex1} we shall consider cases where such an easy algebraic derivation is not possible. To this end, let us first consider the $\mu=5$ case. For this choice, conditions \eqref{exLor} become $t>0$ and $9 \alpha  t^5+t>0$ respectively. As a first step, let us take $\alpha>0$ to simply trivialize the second inequality and have $t$ running in the infinite half-line. In the time gauge that we are, where $Q=-t$, and since \eqref{Qcon1} holds, the Hubble function is just given by $H = \sqrt{\frac{t}{6}}$. In order to obtain its functional behaviour in the cosmic time gauge, where $N=1$, we need to first calculate $\tau$ as a function of $t$ from \eqref{gentimetr}. If we choose the minus sign expression from \eqref{solex1N} (this is done, as we are going to see immediately afterwards, so as to map the function $\tau$ to the positive half-line) and consider that $\alpha$ is positive, then we obtain
\begin{equation}
  \begin{split} \label{ex1time}
    \tau(t) = & \frac{2}{w+1}\sqrt{\frac{2}{3 t}}+ \frac{2^{\frac{3}{4}} \alpha^{\frac{1}{8}}}{3^{\frac{1}{4}} (w+1)} \Bigg\{ \sqrt{\sqrt{2}+1} \Bigg[ \arctan\left(\frac{\left(3-2 \sqrt{2}\right)^{\frac{1}{4}} \left(1-\sqrt{3} \alpha^{\frac{1}{4}}  t\right)}{ 6^{\frac{1}{4}} \alpha^{\frac{1}{8}} \sqrt{t}}\right) \\
     & + \mathrm{arctanh} \left(\frac{6^{\frac{1}{4}} \left(3+2 \sqrt{2}\right)^{\frac{1}{4}} \alpha^{\frac{1}{8}} \sqrt{t}}{\sqrt{3} \alpha^{\frac{1}{4}} t+1}\right) \Bigg] + \sqrt{\sqrt{2}-1} \Bigg[ \arctan\left(\frac{\left(3+2 \sqrt{2}\right)^{\frac{1}{4}} \left(1-\sqrt{3} \alpha^{\frac{1}{4}}  t\right)}{ 6^{\frac{1}{4}} \alpha^{\frac{1}{8}} \sqrt{t}}\right) \\
     &+ \mathrm{arctanh} \left(\frac{6^{\frac{1}{4}} \left(3-2 \sqrt{2}\right)^{\frac{1}{4}} \alpha^{\frac{1}{8}} \sqrt{t}}{\sqrt{3} \alpha^{\frac{1}{4}} t+1}\right)\Bigg] \Bigg\} - C .
  \end{split}
\end{equation}
We already discussed that $t\in (0,+\infty)$, we observe that at the one border value we have $\Lim{t\rightarrow 0^+} \tau(t) = +\infty$, while at the other  $t\rightarrow +\infty$  the limit of $\tau$ equals some finite value. The latter can be  set to zero by choosing appropriately the integration constant $C$. In this case, this particular choice is $C=-\left(\frac{4 \sqrt{2}}{3}+2\right)^{1/4} \pi \alpha^{1/8} \left(1+w\right)^{-1}$. So now, the above function, maps the $t\in (0,+\infty)$ range of the solution \eqref{solex1} to the $\tau \in (0,+\infty)$ in the cosmic time gauge. We can now use the function \eqref{ex1time} of the cosmic time, to make parametric plots of $H(t)=\sqrt{\frac{t}{6}}$ as $H(\tau)$ for various values of the parameters. The process can be repeated for other values of $\mu$ as well. In Figure \ref{fig1} we show the parametric plot of $H(t)$ with respect to $\tau(t)$ for two different $f(Q)$ theories with $\mu=5$ and $\mu=6$, for a radiation matter content, $w=\frac{1}{3}$, and in each graph we also provide the corresponding Hubble function of General Relativity ($\alpha=0$) for comparison. Due to the complexity of the expression, we refrain of giving the $\tau(t)$ for $\mu=6$ here and we just restrict ourselves to presenting the resulting parametric plot. We prefer to present the parametric plots $H(\tau)$ here instead of $H(a)$ since $\tau$ can serve as an ``absolute'' variable for comparison, in contrast to $a(\tau)$ whose evolution with respect to the time $\tau$ changes for different values of the parameters.  Note also, that in the graph for $\mu=6$ we have used a negative $\alpha$ parameter, this is so that the corresponding inequalities \eqref{exLor} are satisfied for $t\in (0,+\infty)$, as for our example of the $\mu=5$ case. We will return to this point a little later in our analysis.

A first observation with respect to Figure \ref{fig1} is that at early times the evolution for different values of $\alpha$ is hardly distinguishable, but as the universe expands a closer to zero value of $\alpha$ tends to the General Relativity solution faster. As far as the $\mu$ value is concerned, we may conclude that as it assumes highest values, it makes the departure from GR more prominent; since, for the same time values, it tends to give higher expansion rates.

\begin{figure}[ptb]
\includegraphics[width=1\textwidth]{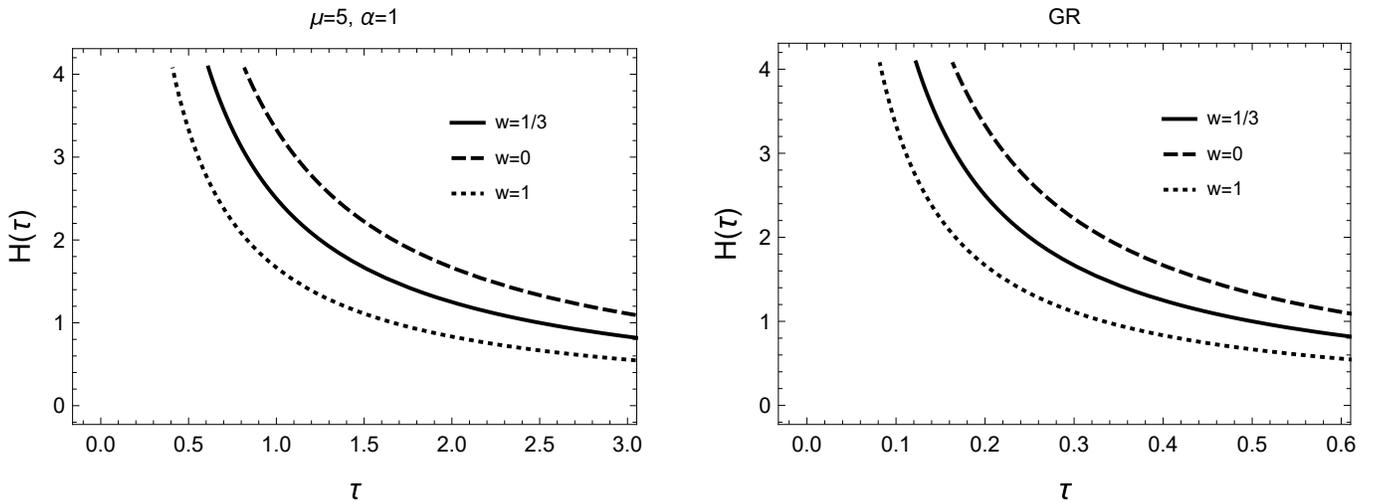}\caption{The behaviour of the Hubble function in $f(Q)=Q+\alpha Q^5$ theory and in General Relativity for various values of the perfect fluid equation of state constant $w$, with respect to the cosmic time $\tau$. The plots are given for $\alpha=1$.}%
\label{fig2}%
\end{figure}

In Figure \ref{fig2}, we explore further the $\mu=5$ case for different values of $w$ (radiation, dust and stiff matter) comparing always next to the relevant results from GR. It seems that the general functional behaviour of the Hubble function is similar in the two theories. However, the $\mu=5$ case gives higher expansions rates for the same values of the cosmic time, $\tau$, variable.

Another interesting observation that we can make is the possibility of obtaining bouncing solutions for appropriate values of the involved parameters. As we previously mentioned, in Fig. \ref{fig1}, we used positive values of $\alpha$ for $\mu=5$, while negative for $\mu=6$. The reason behind this choice is the following: Let us turn to inequalities \eqref{exLor} and consider that $2\mu-1>0$. Then for all odd integers $\mu$ the second inequality is satisfied for $t\in(0,+\infty)$ if $\alpha>0$ and for all even integers $\mu$ if $\alpha<0$. What happens however if we choose $\alpha$ in the opposite manner? The answer is that we get a bounded $t$ in a region $0<t< \frac{1}{(2\mu-1)\alpha}$ for $\mu$ even and $\alpha>0$ and in $0<t< \frac{-1}{(2\mu-1)\alpha}$ for $\mu$ odd and $\alpha<0$. It is for these values of the parameters that a bouncing takes place; albeit a singular one. To facilitate the construction of how this happens, let us consider a simpler model, like $f(Q)=Q + \alpha Q^2$. The $\mu=2$ for this theory is an odd number, so we consider $\alpha>0$. The solution \eqref{solex1} becomes
\begin{subequations} \label{solexbounce}
  \begin{align} \label{solexbounceN}
    N_{\pm} = & \pm \sqrt{\frac{2}{3}} \frac{\sqrt{\frac{2}{3}} (1-6 \alpha  t) (1-2 \alpha  t)^{3/2}}{(w+1) (t (1-2 \alpha  t))^{3/2} (3 \alpha  t-1)} \\ \label{solexbouncea}
    a = & \left(\frac{2\rho_0}{t-3 \alpha  t^2}\right)^{\frac{1}{3( w+1)}} .
  \end{align}
\end{subequations}
As explained before, due to $\alpha$ being positive, and in order for the solution to be Lorentzian, the time variable in this gauge is bound in the region $0<t<\frac{1}{3\alpha}$. Notice however that there is a value in this region for which the lapse function, denoted here by $N_{\pm}$, becomes zero, i.e. $t=\frac{1}{6\alpha}$. This is a possibly problematic point of which we need to take care in the construction of $\tau(t)$. What we will do, is, split the range of the time variable in two parts, one considering $t<\frac{1}{6\alpha}$ and another $t\geq\frac{1}{6\alpha}$, the $t=\frac{1}{6\alpha}$, as we are going to see is a point of a (Riemannian) curvature singularity.

\begin{figure}[ptb]
\includegraphics{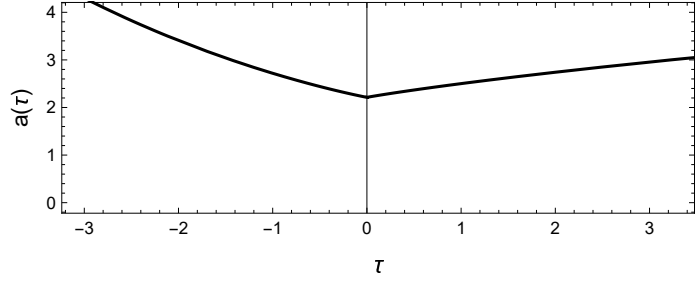}\caption{Parametric plot that depicts the bouncing solution of the scale factor as a function of the comsic time $\tau$, as defined by \eqref{proptbounce}. The graph shows the existing discontinuity in the first derivative. The values that have been used for the involved parameters are $\rho_0=\alpha=1$, $w=\frac{1}{3}$.}%
\label{fig3}%
\end{figure}

According to the previous consideration we define the cosmic time as the following function
\begin{equation} \label{proptbounce}
  \tau(t) = \int\!\! N_{\pm} dt =
         \begin{cases}
           \frac{1}{3(w+1)} \left(\frac{2 \sqrt{6}}{\sqrt{t}} + 6 \sqrt{2} \sqrt{\alpha }\; \mathrm{arctanh}\left(\sqrt{3\alpha t}\right) \right) + C_+, & \mbox{if } 0<t<\frac{1}{6\alpha} \\
           \frac{-1}{3(w+1)} \left(\frac{2 \sqrt{6}}{\sqrt{t}} + 6 \sqrt{2} \sqrt{\alpha } \; \mathrm{arctanh}\left(\sqrt{3\alpha t} \right) \right) + C_- , & \mbox{if } \frac{1}{6\alpha} \leq t<\frac{1}{3\alpha} ,
         \end{cases}
\end{equation}
where for the constants of integration we have $C_\pm = \mp \frac{2\sqrt{\alpha}}{w+1} \left(2+\sqrt{2} \mathrm{arctanh}\left(\frac{1}{\sqrt{2}}\right)\right)$. This last choice has been made so that $\Lim{t\rightarrow\frac{1}{6\alpha}^-}\tau(t)= \Lim{t\rightarrow\frac{1}{6\alpha}^+}\tau(t)=0$. So, we see that the point $t=\frac{1}{6\alpha}$ corresponds to the origin of the cosmic time $\tau=0$. The limits at $t\rightarrow 0$ and $t\rightarrow \frac{1}{3\alpha}$ respectively yield $\tau \rightarrow +\infty$ and $\tau \rightarrow -\infty$. Hence, relation \eqref{proptbounce} defines a continuous one-to-one function that takes values in the whole real line. We notice that in the origin of the cosmic time $\tau=0$ (or equivalently at $t=\frac{1}{6\alpha}$) the scale factor given by \eqref{solexbouncea}, assumes a non-zero minimum value. However, this is a (Riemannian) curvature singularity point as it can be easily seen from the Riemannian Ricci scalar, $\tilde{R} = \frac{t (3 \alpha  t (3 w-5)-3 w+1)}{2(1-6 \alpha  t)}$ corresponding to this solution. We need to also mention that, the scale factor is a continuous function, but its first derivative, i.e. $\frac{da}{d\tau}=\frac{1}{N_{\pm}}\frac{da}{d\tau}$ has a discontinuity at $t=\frac{1}{6\alpha}$. This can be also seen by the graph that we present in Figure \ref{fig3}. The same can also be checked to be true for the second derivative as well.

Thus, for the range of parameters, where $t$ is bounded, we obtain bouncing, but singular solutions; at least from the perspective of the Riemannian scalars, the non-metricity scalar $Q=-t$ is finite since $t$ is bounded. The scale factor is continuous with a minimum non-zero value, but a discontinuity in its derivatives takes place at the origin. This same behaviour can be also derived for the cases $\mu=5$ and $\mu=6$ for the appropriate choices for the range of the parameter $\alpha$ (negative and positive respectively). Thus, we see a pattern forming in $f(Q)= Q + \alpha Q^\mu$ theory with a perfect fluid, where for each value of $\mu$ there exist two types of solutions depending on the sign of $\alpha$: one with a scale factor starting from zero and another where a non-zero value and a bounce can be obtained, which however hides a discontinuity in the derivatives of the metric. We need to mention here that the possibility of bouncing solutions in $f(Q)$ theory has also been investigated in \cite{Myrzakulov}.

\subsubsection{The \texorpdfstring{$f(Q)=Q e^{\frac{q}{Q}}$}{f(Q)=Q exp(q/Q)} example}

Another interesting choice of $f(Q)$ function has been proposed in \cite{Saridakis} and is of the form $f(Q)=Q e^{\frac{q}{Q}}$, where of course, for $q=0$, General Relativity is recovered. We work in a similar fashion to the previous example. The substitution of the $f(Q)$ function under investigation in \eqref{fsol1}, together with the adoption of the time gauge $Q=-t$ leads to the expressions
\begin{subequations} \label{ex2sol}
  \begin{align} \label{ex2solN}
    N= & \pm \sqrt{\frac{2}{3}} \frac{t^2 + q t + 2 q^2}{t^{\frac{5}{2}}\left(w+1\right)\left(t+2 q\right)} \\
    a= & \left(\frac{2 \rho_0 e^{\frac{q}{t}}}{t+2 q}\right)^{\frac{1}{3(1+w)}} .
  \end{align}
\end{subequations}
The corresponding energy density of the fluid in this case is $\rho = \frac{1}{2} e^{-\frac{q}{t}} (2 q+t)$. Following the same reasoning as before, by requiring a positive energy density and a Lorentzian solution, we are led to the restrictions $\rho_0>0$,
\begin{equation} \label{ineqex2}
  t>0 \quad \text{and} \quad t + 2 q >0.
\end{equation}
The latter, lead to two possibilities: one requires just $q>0$ and $t>0$, while the other is $q<0$ and $t>-2q$. The lapse does not become zero at any point in these ranges of values and the procedure followed for both cases is the same.

\begin{figure}[ptb]
\includegraphics{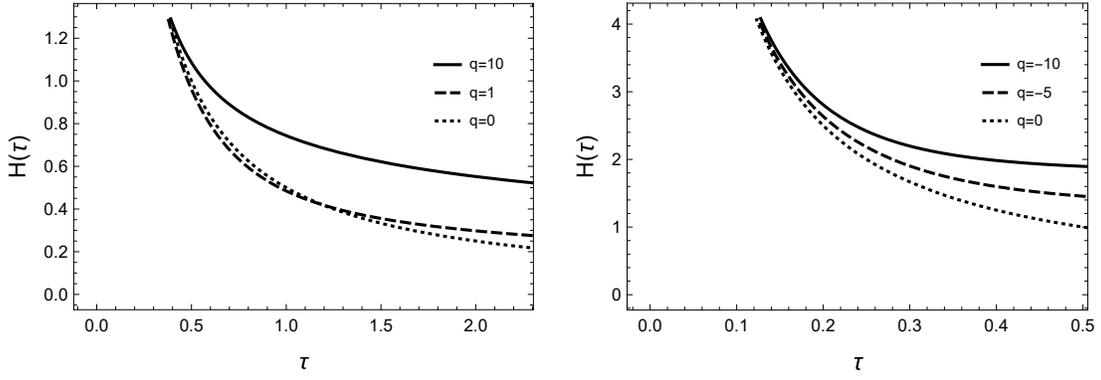}\caption{Parametric plot of the Hubble function with respect to the cosmic time for an equation of state parameter $w=\frac{1}{3}$. The first graph includes positive values of $q$ and the second negative. For comparison, the corresponding GR solution ($q=0$) is given by the dotted line.}%
\label{fig4}%
\end{figure}

For the cosmic time we use again definition \eqref{gentimetr}, which gives
\begin{equation}\label{tauex2}
  \tau(t) = \frac{2}{3 (w+1)}\sqrt{\frac{2}{3}} \left( \frac{q}{t^{3/2}} - \frac{3 \arctan \left(\sqrt{\frac{t}{2q}}\right)}{\sqrt{2 q}} \right) - C.
\end{equation}
For the above expression we used the minus sign of \eqref{ex2solN}, because, once more, it is this what leads to a  $\tau$ ranging from zero to plus infinity. The value of $C$ is set so that the limit of $\tau(t)$ at $t\rightarrow+\infty$, which is finite, becomes zero; thus, we obtain $C= \frac{\pi}{w+1} \sqrt{\frac{1}{3 q}}$. For any of the two cases  we may consider \eqref{ineqex2}, the other limit leads to $\tau\rightarrow +\infty$. For example, in the first case, where $q>0$ and $t>0$, the limit of $t$ going to zero yields $\tau\rightarrow +\infty$, while in the second, where $q<0$ and $t>-2q$,  the limit $t\rightarrow -2q$ yields again $\tau\rightarrow +\infty$. As a result, in both cases, the function \eqref{tauex2} takes values in the positive half-line. The Hubble function in this gauge is of course the same as before, $H=\sqrt{\frac{t}{6}}$. In Figure \ref{fig4}, we give the parametric plot of the latter with respect to $\tau(t)$ for various positive and negative values of $q$.

We notice that the deviation from the GR solution, which is depicted with the dotted line, becomes more important for later times. We additionally observe some differences between the $q>0$ and $q<0$ cases. For the negative values of $q$ we see in general higher expansion rates for the same order of magnitude of the parameter. For example, if we compare the graph for $q=10$ with that of $q=-10$, the latter leads to higher expansion rates for the same $\tau$. Another difference is that for $q>0$, there appears to be a crossing from expansion rates lower of those of GR, for early times, to those higher of GR at later instants of $\tau$, e.g. see how the line for $q=1$ crosses the $q=0$ line of General Relativity at a specific instant of time. The latter does not seem to happen for $q<0$, at least not for the values that are depicted in the graph. At this point we need to mention a limitation regarding these graphs. Due to the fact that the $\tau\rightarrow 0$ corresponds to $t\rightarrow +\infty$, we cannot present plots that go arbitrary close to $\tau = 0$, because that would require giving values to $t$ that go to infinity, which is practically impossible.

\subsection{Second connection}

Here, we assume that the non-zero components of the connection $\Gamma^\mu_{\kappa\lambda}$ are given by \eqref{common} and \eqref{con2}. From the latter set we see that $\gamma(t)$ cannot be zero. Unlike the previous case, this function plays now a dynamical role. The non-metricity scalar reads
\begin{equation}\label{Qcon2}
  Q = - \frac{6 \dot{a}^2}{N^2 a^2}+ \frac{3\gamma}{ N^2} \left(\frac{3\dot{a}}{a} -\frac{\dot{N}}{N} \right) + \frac{3\dot{\gamma}}{N^2}
\end{equation}
and it clearly involves $\gamma$ in its expression. The field equations for the metric result in
\begin{subequations} \label{feq12}
\begin{align} \label{feq12a}
  & \frac{3 \dot{a}^2 f'(Q)}{a^2 N^2} +\frac{1}{2} \left(f(Q)-Q f'(Q)\right) + \frac{3 \gamma \dot{Q} f''(Q)}{2 N^2} = \rho, \\ \label{feq12b}
  & -\frac{2}{N} \frac{d}{dt} \left( \frac{f'(Q) \dot{a}}{N a} \right) - \frac{3 \dot{a}^2}{N^2 a^2} f'(Q) - \frac{1}{2} \left(f(Q)- Q f'(Q)\right) + \frac{3 \gamma \dot{Q} f''(Q)}{2 N^2} = p,
\end{align}
\end{subequations}
while the one for the connection yields
\begin{equation} \label{feq22}
  \dot{Q}^2 f'''(Q)  +  \left[\ddot{Q}  +  \dot{Q} \left( \frac{3 \dot{a} }{a}-\frac{\dot{N} }{N} \right) \right] f''(Q)  =0 .
\end{equation}
In the previous case, we saw that the vacuum solutions become those of General Relativity with a cosmological constant and that the type of $f(Q)$ theory only affected the value of the effective cosmological constant. However, here, due to the dynamical involvement of the connection, we will see that different solutions than the de-Sitter space can emerge in vacuum and that the choice of $f(Q)$ theory makes a difference in the resulting solution space.

So, lets consider the vacuum case $p=\rho=0$ and as a base theory let us choose the function $f(Q)=Q^\mu$, where in the limit $\mu\rightarrow 1$ becomes General relativity. Before proceeding, let us note that for a theory of the form $f(Q)=Q^\mu$, and as long as $\mu>2$, then, any combination of functions $a$, $N$ and $\gamma$ that results in $Q=0$ in \eqref{Qcon2}, is trivially a solution of the equations. Due to the infinity of metrics that satisfy such a relation, we may again consider that this realization does not allow for making specific predictions about the results of the theory, thus, we shall refrain from considering this type of solutions. In fact, the process that we later follow removes completely the possibility of arriving at solutions where $Q$ is a constant altogether.

The constraint equation \eqref{feq12a} can be solved algebraically with respect to $N$. Once more we keep $Q$ as it is and we do not substitute it through expression \eqref{Qcon2}. This is because we are again going to utilize the gauge fixing choice to make $Q$ a particular function of time, which significantly simplifies the resulting equations. Note that, from \eqref{Qcon2}, $Q$ now is not necessarily negative. So, we will just set it this time to be $Q(t)=t$ and  upon the end result we will later see for what domain of definition and for which range of the parameters we may have a solution of Lorentzian signature.

By solving \eqref{feq12a} with respect to $N$ and substituting it inside the equation $Q=t$, where $Q$ is given by \eqref{Qcon2}, we obtain a differential equation that involves the second derivative of the scale factor
\begin{equation} \label{secorda}
  \ddot{a}=\frac{1}{4} \left(\frac{(\mu -1) a^2 \left(t \dot{\gamma}-2 (\mu -1) \gamma\right)}{t^2 \dot{a}}+\frac{2 \dot{a} \left(2 t \dot{\gamma}+3(1-2 \mu ) \gamma\right)}{t \gamma }+\frac{8 (1-2 \mu ) \dot{a}^3}{(\mu -1) a^2 \gamma }+\frac{16 \dot{a}^2}{a}+\frac{6 (\mu -1) a \gamma}{t}\right).
\end{equation}
We use this equation to eliminate $\ddot{a}$ from \eqref{feq12b}, in which we also have substituted the expression for $N$ and set the gauge fixing choice $Q=t$. The result is a simple first order differential equation for $\gamma$
\begin{equation}
  2 (2 \mu -1) t \dot{a}^2+(\mu -1) a^2 \left(2 (\mu -1) \gamma-t \dot{\gamma}\right) =0.
\end{equation}
This can be directly integrated to yield
\begin{equation} \label{solgam}
  \gamma (t) = \frac{2(2\mu-1)t^{2(\mu-1)}}{(\mu-1)}\int\!\! t^{-2(\mu-1)}\frac{\dot{a}^2}{a^2} dt .
\end{equation}
Substitution of the above expression for $\gamma$ into \eqref{secorda} yields an integro-differential equation for the scale factor $a(t)$. However, if we isolate the integral term on one side and take the derivative of that expression, we obtain the following third order equation
\begin{equation} \label{thirdord}
  \dddot{a} = -\frac{8 \dot{a}^3}{a^2}+\frac{2 \left((\mu -2) t \ddot{a} + (\mu -1) \dot{a}\right)}{t^2}+\frac{\dot{a} \left(9 t \ddot{a}-2 (\mu -5) \dot{a}\right)}{t a} .
\end{equation}

The problem of solving the previous equation can be addressed with the help of the theory of symmetries of differential equations. We avoid the technical details here and we refer to well-known textbooks \cite{Olver,Stephani} for more information. We just mention that \eqref{thirdord} admits a three-dimensional algebra of Lie-point symmetry generators, two of which form an Abelian subalgebra. This implies that these can be used to generate a transformation which will both reduce the order of the equation and also make it autonomous. In our case this transformation is
\begin{equation}\label{transf}
  a = e^{\frac{1}{6} (1-2 \mu ) \omega(s)+s}, \quad t = e^{\omega(s)},
\end{equation}
where $\omega(s)$ and $s$ are the new dependent and independent variables that are going to substitute $a(t)$ and $t$. Under the change of variables implied by \eqref{transf}, equation \eqref{thirdord} becomes
\begin{equation}\label{tranthird}
  3 \left(\frac{d^2\omega}{ds^2}\right)^2+ \frac{d\omega}{ds} \left(6 \frac{d^2\omega}{ds^2}-\frac{d^3\omega}{ds^3}\right) =0 .
\end{equation}
As we see, the new equation truly is autonomous, since now there is no explicit dependence in $s$, and it is also effectively a second order differential equation because no $\omega(s)$ term appears. The general solution of \eqref{tranthird} is
\begin{equation} \label{solthird}
  \omega(s) = \lambda_3 + \lambda_2 \ln\left( \frac{1-\sqrt{A(s)} \sqrt{A(s)^2-3 A(s)+3}}{1+\sqrt{A(s)} \sqrt{A(s)^2-3 A(s)+3}} \right), \quad \text{where} \quad A(s) = 1 - \lambda_1 e^s
\end{equation}
and $\lambda_i$, $i=1,2,3$, are constants of integration.

Of course, at this point, we need use expression \eqref{solthird} in \eqref{transf} to obtain the resulting $a(t)$, which, together with the $\gamma(t)$ of \eqref{solgam}, are to be substituted in the original equations to see under which conditions they form a solution. This is necessary since we did not solve the actual equations, but another of higher order, so we expect to have in our expressions at least one redundant constant of integration. Through this process, we conclude that the constant of integration that results from the calculation of the integral in \eqref{solgam} needs to be set equal to zero. The desired triplet that we finally obtain is
\begin{subequations} \label{sol2}
  \begin{align} \label{sol2N}
    N(t) & = \pm \frac{\sqrt{\frac{2}{3}} \sqrt{\kappa } \lambda  t^{\frac{\lambda -3}{2}}}{\sqrt{\frac{\mu -1}{\mu }} \left(\kappa -t^{\lambda } \right)} , \\ \label{sol2a}
    a(t) & = \frac{a_0 t^{\frac{1}{6} (\lambda -2 \mu +1)}}{\left(\kappa -t^{\lambda } \right)^{\frac{1}{3}}}, \\
    \gamma(t) & = \frac{\kappa  (\lambda -2 \mu +1)^2-(\lambda +2 \mu -1)^2 t^{\lambda }}{18 (\mu -1) t \left(t^{\lambda }-\kappa \right)},
  \end{align}
\end{subequations}
where, in order to simplify the expressions, we have made the re-parameterizations of the original constants of integration as $\lambda_2=\frac{1}{\lambda}$ and $\lambda_3= \frac{\ln(-\kappa)}{\lambda}$. The constant $\lambda_1$, together with the rest of the constants that appear multiplicatively in the expression for $a(t)$ can be normalized to any value through a constant scaling of the radial variable $r$ in the line element. We choose to depict this arbitrariness with the constant $a_0$ appearing in \eqref{sol2a}. As it is evident from \eqref{sol2}, the case $\mu=1$ is excluded from this solution. This has to do with our gauge fixing choice of $Q=t$, which straightforwardly excludes the GR vacuum solution, where $Q=$constant.

It can be directly checked, that relations \eqref{sol2}, not only solve the set of equations \eqref{feq12} and \eqref{feq22} for $f(Q)=Q^\mu$, but also, upon substitution in \eqref{Qcon2}, yield $Q=t$, which verifies the consistency of the result. Thus, we obtain the general solution for $f(Q)=Q^\mu$ theory in the time gauge where $Q=t$ is the time parameter. Of course, going to the cosmic time gauge, where $N(\tau)=1$ will not be possible for every value of the parameters, since the inverse of the transformation in general will not be expressed in terms of elementary functions. For example, by using \eqref{sol2N} we see that, for $\lambda\neq 1$ and $\frac{3\lambda-1}{2\lambda} \notin \mathbb{Z}_{-}\cup \{0\}$,
\begin{equation} \label{2ndconpropt}
  \int N(t) dt = \tau + C \Rightarrow \pm \frac{2 \sqrt{\frac{2}{3}} \lambda  t^{\frac{\lambda -1}{2}}}{\sqrt{\kappa } (\lambda -1) \sqrt{\frac{\mu -1}{\mu }}} { }_2F_1\left(1,\frac{\lambda -1}{2 \lambda };\frac{3\lambda-1}{2\lambda};\frac{t^{\lambda }}{\kappa }\right) = \tau + C,
\end{equation}
where ${ }_2F_1$ is the Gauss hypergeometric function. The values for which the integral of \eqref{sol2N} can be calculated in terms of elementary functions are specified by Chebyshev's theorem \cite{Rittbook}, which is used in cosmology to distinguish analytic solutions of the Friedmann equations \cite{Gibbons,Faraoni}. In order to comply with the requirements of the theorem, $\lambda$ needs to be just any rational number, i.e. $\lambda \in \mathbb{Q}$. Of course this still does not guarantee that the expression will be easily inverted to obtain $t(\tau)$.  Acquiring the inverse, $t(\tau)$, is achievable for very specific values of the parameter $\lambda$. We can use however the $\tau(t)$, as previously, to make parametric plots of the Hubble function $H(t)$ and get a glimpse of the time evolution in the cosmic time gauge for any $\lambda$.

\begin{figure}[ptb]
\includegraphics{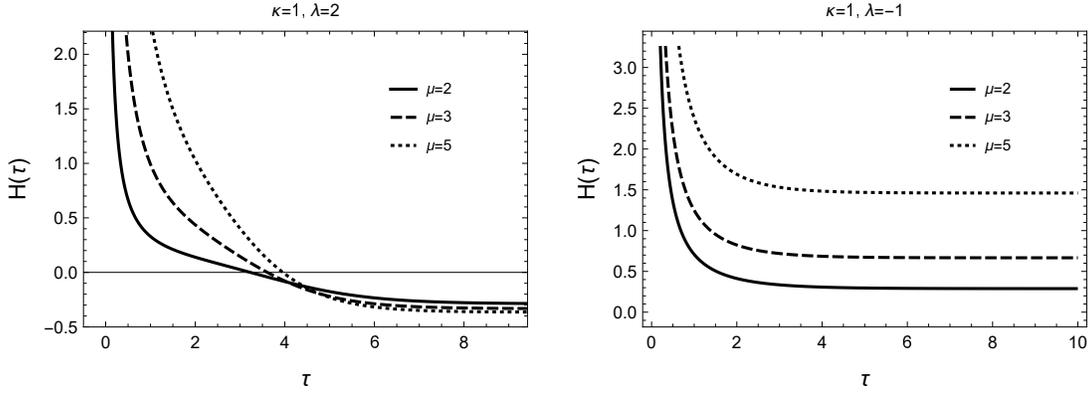}\caption{Graphs of the Hubble function with respect to the cosmic time for two different pairs of $\kappa$, $\lambda$ and for various different values of the power $\mu$ of $f(Q)=Q^\mu$.}%
\label{fig5}%
\end{figure}

For the derived solution \eqref{sol2}, the Hubble function is (in the $t$ time variable)
\begin{equation}
  H(t)= \frac{1}{Na} \frac{da}{dt} = \frac{1}{2 \lambda } \sqrt{\frac{\mu -1}{6 \kappa  \mu }} \left[ \kappa  (2 \mu -\lambda  -1)+(1 -\lambda -2 \mu ) t^{\lambda } \right] t^{\frac{1-\lambda}{2}} .
\end{equation}
In Figure \ref{fig5} we present two plots of $H(t)$ with respect to $\tau(t)$ for two different sets of values of $\kappa$, $\lambda$ and for various values of $\mu$. Interestingly enough, the functional behaviour for the integration constants, beside $\mu$, can be quite different. The first graph depicts the case $\kappa=1$, $\lambda=2$, where for the calculation of $\tau$ we used the positive sign branch of \eqref{sol2N}. The latter is a real function that takes values in the range $\tau \in (0,+\infty)$ as long as $t\in (0,1)$ (the constant $C$ is set to zero). Note, that having $ 0< t<1$, with $\kappa=1$ and $\lambda=2$, does not really cause a problem in \eqref{sol2a}, in the sense of the latter taking imaginary values, because there is the arbitrary constant $a_0$ in the solution, inside which, we can ``absorb'' any complex constant number. As we see in the first plot of Fig. \ref{fig5} this solution yields a universe which initially  expands, with a continuously diminishing expansion rate. Then, after a finite time, a contraction phase takes place. On the other hand, we obtain a completely different behaviour by setting $\lambda=-1$. In fact, for this value we can invert the expression for the cosmic time. For $\kappa=1$, $\lambda=-1$, the integral yielding the cosmic time is simply
\begin{equation} \label{2ndconpropt2}
  \tau(t)=\int N(t) dt = \pm \frac{\sqrt{\frac{2}{3}} \log \left(\frac{t-1}{t}\right)}{\sqrt{\frac{\mu -1}{\mu }}} -C.
\end{equation}
We can eliminate the constant of integration $C$ and invert the above relation to get directly the function $H(\tau)$, which is (considering the positive branch of \eqref{2ndconpropt2})
\begin{equation}
  H(\tau)= \left(\frac{\mu -1}{6 \mu }\right)^{\frac{1}{2}} \frac{(\mu -1) e^{\sqrt{\frac{3}{2}} \sqrt{\frac{\mu -1}{\mu }} \tau}+1}{e^{\sqrt{\frac{3}{2}} \sqrt{\frac{\mu -1}{\mu }} \tau}-1} .
\end{equation}
The plot of this function is seen in the second part of Fig. \ref{fig5} and depicts an expanding universe with an ever slower expansion rate as time progresses. Unlike the previous case however, it never results in a contracting phase.

\subsection{Third connection}

Here, we make use of the connection with non-zero components given by \eqref{common} together with \eqref{con3}. This connection is also involved in the dynamics of the system. The non-metricity scalar assumes the form
\begin{equation}\label{Qcon3}
  Q = - \frac{6 \dot{a}^2}{N^2 a^2}+ \frac{3\gamma}{ a^2} \left(\frac{\dot{a}}{a} +\frac{\dot{N}}{N} \right) + \frac{3\dot{\gamma}}{a^2} .
\end{equation}
The equations of motion for the metric are
\begin{subequations}\label{feq13}
  \begin{align}\label{feq13a}
  & \frac{3 \dot{a}^2 f'(Q)}{a^2 N^2} +\frac{1}{2} \left(f(Q)-Q f'(Q)\right) - \frac{3 \gamma \dot{Q} f''(Q)}{2 a^2} = \rho, \\ \label{feq13b}
  & -\frac{2}{N} \frac{d}{dt} \left( \frac{f'(Q) \dot{a}}{N a} \right) - \frac{3 \dot{a}^2}{N^2 a^2} f'(Q) - \frac{1}{2} \left(f(Q)- Q f'(Q)\right) + \frac{ \gamma \dot{Q} f''(Q)}{2 a^2} = p
  \end{align}
\end{subequations}
and for the connection
\begin{equation} \label{feq23}
  \dot{Q}^2 f'''(Q)  +  \left[\ddot{Q}  +  \dot{Q} \left(\frac{\dot{a}}{a} + \frac{\dot{N}}{N}+  \frac{2 \dot{\gamma}}{\gamma} \right) \right] f''(Q)  =0 .
\end{equation}
Once more, we shall consider the vacuum case $p=\rho=0$ in the context of a $f(Q)=Q^\mu$ theory. We employ a similar strategy as before, but with a few modifications. First, we solve the constraint equation \eqref{feq13a} for the lapse $N(t)$. We substitute this result into equation \eqref{feq13b} and make the gauge fixing choice $Q=t$, the result being the following equation which is algebraic in $a(t)$:
\begin{equation}
  (2 \mu -1) t^2 a^2+3 \mu  \left(t \dot{\gamma}+(2 \mu -3) \gamma\right) =0.
\end{equation}
Assuming that $\mu\neq \frac{1}{2}$ we may solve for $a$ and substitute this result, together with all the previous assertions, in the only remaining equation \eqref{feq23}. Subsequently, we arrive at the following third order equation for $\gamma$
\begin{equation} \label{gamthird}
  \begin{split}
    & 2 t^3 \gamma \left(t \dot{\gamma}-2 \gamma \right) \dddot{\gamma} - t^4 \gamma  \ddot{\gamma}^2 + 4 t^2  \left(t^2 \dot{\gamma}^2 + (\mu -4) t \gamma \dot{\gamma} + (5-2 \mu ) \gamma^2\right) \ddot{\gamma} \\
    & + 8 (\mu -2) t^3 \dot{\gamma}^3+4 ((\mu -15) \mu +23) t^2 \gamma  \dot{\gamma}^2-8 (\mu  (2 \mu -17)+22) t \gamma^2 \dot{\gamma} +4 (2 \mu -9) (2 \mu -3) \gamma^3 = 0.
  \end{split}
\end{equation}
The above equation is rather tedious, however there do exists some symmetries that allow for its simplification. If we perform the transformation
\begin{equation}
  \gamma = \exp\left[\int\!\!(1+2 s)\omega(s)ds\right], \quad t = \exp\left(\int\!\! s\omega(s)ds\right),
\end{equation}
with $s$, $\omega(s)$ the new independent and dependent variables respectively, then the equation is reduced to first order Abel equation
\begin{equation}
  2 s \frac{d\omega}{ds}+ s^2 \left(\left(4 \mu ^2-1\right) s^2+4 (3 \mu -1) s+5\right) \omega^3-4 s (\mu  s+2) \omega^2+5 \omega =0.
\end{equation}
Unfortunately, we did not manage to associate the later to some known integrable class. However, we do have to report an exact particular solution of the original equation \eqref{gamthird}, which is in the form of a power law with respect to $t$.  The solving triplet is
\begin{subequations} \label{sol3}
  \begin{align}
    N(t) & = \pm \left[\frac{3 \mu  (4 \mu -3)}{(2 \mu +1) (1-\mu )}\right]^{\frac{1}{2}} \frac{2 \mu +1}{5 t^{3/2}} \\
    a(t) & = \sigma \left[\frac{6 \mu  (3-4 \mu )}{5 (2 \mu -1)}\right]^{\frac{1}{2}} t^{-\frac{1}{10} (2 \mu +1)} \\
    \gamma(t) & = \sigma^2 t^\frac{9-2\mu}{5} ,
  \end{align}
\end{subequations}
which it is easily seen that it is compatible with our gauge fixing choice $Q=t$. However, there are certain limits constraining the parameters in order to have a Lorentzian signature of the metric. One choice is that both the expressions inside the square roots appearing in $N$ and $a$ have to be positive; this leads to $-\frac{1}{2}<\mu<0$. The other possibility is to have only the expression appearing inside the square root in $N$ positive and the one in $a$ being negative, then the imaginary unit appearing in $a(t)$ can be absorbed inside $\sigma$, by considering the latter to be a purely imaginary number, this leads to the restriction $\frac{3}{4}<\mu<1$ with $\sigma \in \mathrm{i}\mathbb{R}-\{0\}$. These restrictions are however with the condition that $t$ is positive in \eqref{sol3}. Since, it is quite convenient with working with a positive $t$, the easiest way to take a Lorentzian solution for the other values of the parameter $\mu$ is to go back and instead of fixing the gauge to $Q=t$, fix it as $Q=-t$ (or equivalently making a change $t\rightarrow -t$ in \eqref{sol3}, with an appropriate reparametrization of $\sigma$). The end result can be written as
\begin{subequations} \label{sol3sec}
  \begin{align}
    N(t) & = \pm \left[\frac{3 \mu  (3-4 \mu)}{(2 \mu +1) (1-\mu )}\right]^{\frac{1}{2}} \frac{2 \mu +1}{5 t^{3/2}} \\
    a(t) & = \sigma \left[\frac{6 \mu  (4 \mu - 3 )}{5 (2 \mu -1)}\right]^{\frac{1}{2}} t^{-\frac{1}{10} (2 \mu +1)} \\
    \gamma(t) & = \sigma^2 t^\frac{9-2\mu}{5} ,
  \end{align}
\end{subequations}
and it now yields $Q=-t$. For a positive $t$, there are again two possibilities for a Lorentzian metric. The first is, like before, to have both expressions under the square roots being positive. This leads to $0<\mu<\frac{1}{2}$ or $\mu>1$. The second option yields $\mu<-\frac{1}{2}$ or $\frac{1}{2}<\mu<\frac{3}{4}$ with the necessary supplementary condition $\sigma \in \mathrm{i}\mathbb{R}-\{0\}$.

Both of the above solutions can be easily transformed into the cosmic time gauge. We just need to use \eqref{gentimetr} to derive the $t(\tau)$ relation, then the scalars $a(t)$ and $Q(t)$ can be easily calculated by a straightforward substitution. For the $\gamma(t)$ however, we need to remind ourselves that this is not a scalar, so we cannot simply substitute in it the $t(\tau)$. Its transformation law can be derived from the general transformation law of a connection
\begin{equation}
  \bar{\Gamma}^{\lambda}_{\;\mu\nu} = \frac{\partial \tilde{x}^\lambda}{\partial x^\rho} \frac{\partial x^\eta}{\partial \tilde{x}^\mu} \frac{\partial x^\sigma}{\partial \tilde{x}^\nu} \Gamma^\rho_{\;\eta\sigma} - \frac{\partial x^\rho}{\partial \tilde{x}^\nu} \frac{\partial x^\sigma}{\partial \tilde{x}^\mu} \frac{\partial^2 \tilde{x}^\lambda}{\partial x^\rho \partial x^\sigma},
\end{equation}
which in our case results in $\gamma(\tau) = \gamma(t(\tau))\left(\frac{dt(\tau)}{d\tau}\right)^{-1}$. With this in consideration, after making the appropriate calculations, both of the previous sets can be mapped to the following expressions for $a$ and $\gamma$ in the gauge $N(\tau)=1$
\begin{subequations} \label{sol3cosmic}
  \begin{align} \label{sol3cosmica}
    a(\tau) & = \bar{\sigma} \left[\frac{5 (\mu -1) (4 \mu -3)}{(2 \mu -1) (2 \mu +1) (3-4 \mu )}\right]^{\frac{1}{2}} \tau^{\frac{1}{5} (2 \mu +1)}, \\
    \gamma(\tau) & = \bar{\sigma} \tau^{\frac{1}{5} (4 \mu -3)},
  \end{align}
\end{subequations}
where $\bar{\sigma}$ is a new constant that we introduce to simplify the product of the multiplying constants that appears in the expression of $\gamma(\tau)$ after the transformation. The $\bar{\sigma}$ can be either real or imaginary, depending on the sign of the expression inside the square root of \eqref{sol3cosmica}, so that the $a^2$ remains positive and the signature in the metric is Lorentzian. The ensuing non-metricity scalar, $Q(\tau)$, is given by
\begin{equation} \label{cosmicQ3}
  Q(\tau) = \frac{12 \mu  (2 \mu +1) (3-4 \mu )}{25 (\mu -1) \tau^2}.
\end{equation}
It is straightforward to verify that \eqref{sol3cosmic} and \eqref{cosmicQ3} solve the equations for $f(Q)=Q^\mu$ in the cosmic time gauge $N=1$.

The fact that we get a power-law solution for the scale factor in this time gauge is reminiscent of the solution that one gets in General Relativity in the presence of a perfect fluid, which is characterized by a linear barotropic equation. Truly, if we calculate the effective energy-momentum tensor, as defined by \eqref{Teff}, and consider $\mathcal{T}^{\mu}_{\phantom{\mu}\nu}= \mathrm{diag}(-\rho_{\text{eff}},p_{\text{eff}},p_{\text{eff}},p_{\text{eff}})$, then we see that for \eqref{sol3cosmic} and \eqref{cosmicQ3} we get
\begin{equation}
  p_{\text{eff}} = \frac{(2 \mu +1) (7-6 \mu )}{25 \tau^2}, \quad \rho_{\text{eff}} =  \frac{3 (2 \mu +1)^2}{25 \tau^2}
\end{equation}
with an effective equation of state parameter
\begin{equation}
  w_{\text{eff}} =\frac{p_{\text{eff}}}{\rho_{\text{eff}}}= \frac{7-6 \mu }{3(2 \mu +1)} .
\end{equation}
The phantom divide line $w_{\text{eff}}=-1$ is set at $\mu\rightarrow\pm\infty$. There is also a critical value $\mu=-\frac{1}{2}$, which is excluded by the solution, since it appears also in the denominator of the scale factor \eqref{sol3cosmica}. We notice that theories with $\mu>-\frac{1}{2}$ have $w_{\text{eff}}>-1$, while those of  $\mu<-\frac{1}{2}$ correspond to $w_{\text{eff}}<-1$.

We see how, in this case, the non-trivial connection affects the dynamics. We obtain a vacuum solution, which has the same effect as that of a perfect fluid energy momentum tensor in General Relativity. Only that in this case the effective fluid contribution is related to the geometry and the non-zero connection. Of course, since it is a flat connection, a coordinate system can be found in which $\Gamma^{\lambda}_{\;\mu\nu}$ becomes zero. However, such a transformation would also change the FLRW metric introducing non-diagonal terms. Thus, we see in practice that assuming $\Gamma^{\lambda}_{\mu\nu}=0$ in the same coordinate system where the homogeneity and isotropy of the FLRW metric is obvious, is not a necessity. There exist admissible non-zero connections in the coordinate system where the metric is given by \eqref{genlineel}, which lead to distinct solutions and they affect the dynamics.

\section{Spatially curved models} \label{sec5}

For a non-zero spatial curvature $k$, the connection \eqref{conk1} is to be used. Now, the non-metricity scalar acquires the form
\begin{equation}\label{Qconk}
  Q = - \frac{6 \dot{a}^2}{N^2 a^2}+ \frac{3\gamma}{ a^2} \left(\frac{\dot{a}}{a} +\frac{\dot{N}}{N} \right) + \frac{3\dot{\gamma}}{a^2} + k \left[\frac{6}{a^2} + \frac{3}{\gamma N^2 } \left(\frac{\dot{N}}{N} + \frac{\dot{\gamma}}{\gamma} -\frac{3\dot{a}}{a} \right)\right].
\end{equation}
We see that upon setting $k=0$, it reduces to that corresponding to the third connection of the $k=0$ case, as obtained in \eqref{Qcon3}. The equations of motion for the metric yield
\begin{subequations}\label{feq1k}
  \begin{align}\label{feq1ka}
  & \frac{3 \dot{a}^2 f'(Q)}{a^2 N^2} +\frac{1}{2} \left(f(Q)-Q f'(Q)\right) - \frac{3 \gamma \dot{Q} f''(Q)}{2 a^2} + 3 k \left(\frac{ f'(Q)}{a^2}-\frac{ \dot{Q} f''(Q)}{2 \gamma N^2 }\right) = \rho, \\ \label{feq1kb}
  & -\frac{2}{N} \frac{d}{dt} \left( \frac{f'(Q) \dot{a}}{N a} \right) - \frac{3 \dot{a}^2}{N^2 a^2} f'(Q) - \frac{1}{2} \left(f(Q)- Q f'(Q)\right) + \frac{ \gamma \dot{Q} f''(Q)}{2 a^2} - k \left(\frac{f'(Q)}{a^2}+\frac{3 \dot{Q} f''(Q)}{2 \gamma N^2 }\right) = p ,
  \end{align}
\end{subequations}
and the field equation for the connection becomes
\begin{equation} \label{feq2k}
  \dot{Q}^2 f'''(Q) \left(1+\frac{k  a^2}{N^2 \gamma^2}\right) +  \left[\ddot{Q} \left(1+ \frac{k a^2}{N^2 \gamma^2}\right) +  \dot{Q} \left(\left(1+ \frac{3 k a^2}{N^2 \gamma^2}\right)\frac{\dot{a}}{a} + \left(1-\frac{k a^2}{N^2 \gamma^2}\right) \frac{\dot{N}}{N}+  \frac{2 \dot{\gamma}}{\gamma} \right) \right] f''(Q)  =0 .
\end{equation}
The situation with these equations is quite more complicated and the same trick we performed previously by adopting a gauge fixing that utilized $Q$ (or $-Q$) as the time parameter is not so helpful. However, we are able to disclose an exact solution in the case of a vacuum $\rho=p=0$, $f(Q)=Q^\mu$, theory, to which we arrive in the manner that we subsequently describe. Of course, as previously stated, one can disclose infinitely many solutions by enforcing $Q=0$ and $\mu>2$, but as we explained, we are not interested in obtaining this type of solutions.

In order to proceed, it is first useful to remember that most cosmological solutions that we know from General Relativity, when $k\neq 0$, are expressible in terms of elementary functions in the conformal time gauge, i.e. when $N=a$. So, by making now this gauge fixing choice and additionally enforcing the restrictive condition, that the function $\gamma$ is equal to a particular constant, namely $\gamma = \mp \sqrt{-k}$, then we observe that the constraint equation \eqref{feq1ka} with the substitution of $Q$ from \eqref{Qconk} is easily integrated to give
\begin{equation}\label{solk1}
  a(t) = a_0 e^{\pm \frac{\sqrt{-k} t}{2\mu -1}},
\end{equation}
where $a_0$ is a constant which we can normalize to unity through a combined scaling transformation in $t$, $r$ and $k$. The conditions we have set, together with \eqref{solk1}, satisfy all equations \eqref{feq1k} and \eqref{feq2k}. Of course the solution is real only for a negative spatial curvature $k=-1$. In solution \eqref{solk1} we recognize a Milne-like universe. It can be easily seen that $a(\tau)\propto \tau$ in the gauge $N(\tau)=1$. However, the difference is in the parameter $\mu$ of the theory. The solution corresponds to a Riemann flat universe, when $\mu=1$, which is the Milne case. The non-metricity scalar of the solution is given by
\begin{equation} \label{Qwithkspsol}
  Q= \frac{24 k \mu ^2  }{(1-2 \mu )^2 a_0^2} e^{\mp \frac{2 \sqrt{-k} t}{2 \mu -1}}
\end{equation}
and it does not become a constant even at the limit where the solution of General Relativity is recovered ($\mu=1$). We need to note however, that the $\gamma = \mp \sqrt{-k}$ condition for the function appearing in the connection is not necessary when $\mu=1$. In the latter case, any arbitrary function $\gamma(t)$ serves, together with $N=a=\exp\left(\pm\sqrt{-k} t\right)$. This arbitrariness is also carried in the value of $Q$; for, $\mu=1$, the latter reads
\begin{equation}
  Q = \frac{3 e^{\mp 2 \sqrt{-k} t }}{a_0^2 \gamma} \left[ \left(k +\gamma^2\right)\dot{\gamma} \pm 2 \sqrt{-k} \gamma^3 + 4 k \gamma^2 \pm 2 (-k)^{\frac{3}{2}} \gamma \right]
\end{equation}
in which $\gamma$ remains a free function. Substitution of $\gamma=\mp \sqrt{-k}$, results of course in the expression \eqref{Qwithkspsol} with $\mu=1$. However, as we stated, these values of $\gamma$ become necessary for the satisfaction of the field equations only when $\mu\neq1$.

\section{Conclusion} \label{sec6}

We studied the effect that different connections have in the dynamics of FLRW cosmology in the context of $f(Q)$ theory. The spatially flat case admits three different families of connections. The first, corresponds to the most studied case in the literature, of the coincident gauge. We managed, by using $Q$ as the time variable of the problem, to express the general solution of a perfect fluid for an arbitrary $f(Q)$ theory. The final solution involves an arbitrary function in the connection, which does not affect the gravitational equations. We need to mention however, that it is not clear if this degeneracy can affect the motion of a particle in such a spacetime. In Riemannian geometry the auto-parallel and the extremal length curves coincide, but in theories with non-metricity this is not the case \cite{geo1,geo2}. If we are to assume that the geodesic equations are still given with respect to the metric compatible Levi-Civita connection (see the appendix of \cite{geotrin} for the interpretation of non-metricity in these equations), then the arbitrariness of the function $\gamma(t)$ should not affect them.

The other two connections of the spatially flat model offer quite more complicated dynamics. The existence and the derivation of solutions for these connections is rarely encountered in the literature. This is because most authors assume directly the coincident gauge in a FLRW universe in Cartesian coordinates, which is the dynamically equivalent of the first connection we studied. However, we do see how distinct it is to assume these two connections in place of the first.  Their equation of motion is not satisfied identically and both of them are also involved in the definition of $Q$. By using the same trick as in the first connection, namely choosing the non-metricity as the time variable, we managed to extract new solutions. For the second connection, we derived the general vacuum solution for a power-law $f(Q)$ theory, while for the third we were restricted to just obtaining a partial exact solution. However, we did manage to reduce the problem up to the integration of an Abel equation.

We need to stress that the choice of the non-metricity scalar $Q$ as the time variable of the system, apart from simplifying the equations, served to also guarantee that we would acquire solutions that go outside the scope of General Relativity. This, because the condition $Q\neq$const. automatically excludes the possibility of the theory becoming dynamically equivalent to General Relativity with a cosmological constant. This is what we exactly wanted to study with this work; investigate the possibilities of going beyond General Relativistic solutions. From the particular examples, which we studied in the flat case, it is clear that the theory has rich dynamics and can give various interesting behaviours; from bouncing solutions to inflationary expansions, and even reproducing power-law GR-type of perfect fluid solutions in the absence of matter.

The spatially non-flat metric, leads to severely more complicated equations. Surprisingly enough, in the presence of non-metricity, and for a power-law $f(Q)$ function, we derived a special solution which is reminiscent of the Milne solution in GR. For the future we plan to expand this study by including various types of matter for all possible cases of the admissible connections.

\begin{acknowledgments}
  N. D. acknowledges the support of the Fundamental Research Funds for the Central Universities, Sichuan University Full-time Postdoctoral Research and Development Fund No. 2021SCU12117
\end{acknowledgments}

\end{document}